\documentclass[11pt]{article}
\usepackage{latexsym,amsmath,amssymb,pstricks,enumerate,
epsfig}
\usepackage{amsthm}
\usepackage{mathrsfs}

\advance\textheight by \topskip
\begin{document}
\baselineskip 18pt
\parindent 16pt

\linespread{1.3}

\vfuzz2pt 
\hfuzz2pt 

\newcommand{\be}{\begin{equation}}
\newcommand{\ee}{\end{equation}}
\newcommand{\beu}{\begin{equation*}}
\newcommand{\eeu}{\end{equation*}}
\newcommand{\bea}{\begin{eqnarray}}
\newcommand{\eea}{\end{eqnarray}}
\newcommand{\beaa}{\begin{eqnarray*}}
\newcommand{\eeaa}{\end{eqnarray*}}
\newcommand{\bmx}{\begin{pmatrix}}
\newcommand{\emx}{\end{pmatrix}}

\newcommand{\ol}{\overline}
\newcommand{\ul}{\underline}
\newcommand{\del}{\nabla}
\newcommand{\liealg}[1]{\mathfrak{#1}}
\newcommand{\gC}{\mathfrak g ^ \mathbb C}
\newcommand{\h}{{\mathfrak h}}
\newcommand{\f}{{\mathfrak f}}
\newcommand{\q}{{\mathfrak q}}
\newcommand{\dd}{{\frak d}}
\newcommand{\tD}{{\widetilde D}}
\newcommand{\vv}{{\bf v}}
\newcommand{\vn}{{\bf n}}
\newcommand{\am}{{\alpha}}
\newcommand{\gm}{{\gamma}}
\newcommand{\A}{{\mathsf A}}
\newcommand{\B}{{\mathsf B}}
\newcommand{\hs}{{*}}
\newcommand{\ad}{\,{\rm ad}\,}
\newcommand{\p}{\,{\rm P}}

\newcommand{\ah}{{\hat \alpha}}
\newcommand{\bh}{{\hat \beta}}
\newcommand{\gh}{{\hat \gamma}}
\newcommand{\deh}{{\hat \delta}}
\newcommand{\dem}{{\delta}}
\newcommand{\eh}{{\hat \epsilon}}
\newcommand{\eem}{{\epsilon}}
\newcommand{\roh}{{\hat \rho}}
\newcommand{\kh}{{\hat \kappa}}
\newcommand{\km}{{\kappa}}
\newcommand{\mh}{{\hat \mu}}
\newcommand{\lh}{{\hat \lambda}}
\newcommand{\J}{\,\mathbb J}
\newcommand{\D}{{\cal D}}
\newcommand{\dxy}{\delta(x{-}y)}
\newcommand{\dpxy}{\delta'(x{-}y)}
\newcommand{\half}{\frac{1}{2}}
        \newcommand{\threehalf}{{\textstyle{3\over2}}}
\newcommand{\quarter}{{\textstyle{1\over4}}}
\newcommand{\nn}{\nonumber}
\newcommand{\sign}{{\rm sign}}
\newcommand{\X}{{\tilde X}}
\newcommand{\8}{{\infty}}
\newcommand{\psl}{{\mathfrak{psl}}}

\newcommand{\tr}{\,{\rm tr}\,}
\newcommand{\trr}[1]{\,{\rm tr}|_{#1}\,}
\newcommand{\trl}{\,{\rm tr}|_l\,}
\newcommand{\str}{{\,\rm str\,}}
\newcommand{\rank}{{\rm rank}}
\newcommand{\pf}{{\rm Pf}}
\newcommand{\hl}{\\ \hline \\}
\newcommand{\Ad}{{\rm Ad}}
\newcommand{\orb}{{\rm orb}}
\newcommand{\stab}{{\rm stab}}
\newcommand{\hc}{{\dagger}}
\newcommand{\qbar}{{\bar q}}
\newcommand{\kabar}{{\bar \kappa}}
\newcommand{\betab}{{\beta^B}}

\theoremstyle{definition}
\newtheorem{thm}{Theorem}[section]
\newtheorem{cor}[thm]{Corollary}
\newtheorem{lem}[thm]{Lemma}
\newtheorem{prop}[thm]{Proposition}
\theoremstyle{definition}
\newtheorem{defn}[thm]{Definition}
\theoremstyle{definition}
\newtheorem{fact}[thm]{Fact}
\theoremstyle{definition}
\newtheorem{rem}[thm]{Remark}
\numberwithin{equation}{section}


\newcommand{\Z}{\mathbb{Z}}
\newcommand{\R}{\mathbb{R}}
\newcommand{\Pb}{\mathbf{P}}
\newcommand{\Tb}{\mathbf{T}}
\newcommand{\TwoMink}{\mathbb{R}^{1,1}}
\newcommand{\TwoHMink}{\mathbb{H}^{1,1}}
\newcommand{\Ppoint}{(\mathscr{P})}
\newcommand{\DMink}{\mathbb{R}^{1,D-1}}
\newcommand{\SOC}{SO^\uparrow}

\newcommand{\so}{\mathfrak{so}}
\newcommand{\su}{\mathfrak{su}}
\newcommand{\iso}{\mathfrak{iso}}
\newcommand{\kfrak}{\mathfrak{k}}
\newcommand{\pfrak}{\mathfrak{p}}
\newcommand{\hfrak}{\mathfrak{h}}

\newcommand{\gab}{g_{\alpha\beta}}
\newcommand{\gabu}{g^{\alpha\beta}}
\newcommand{\Bab}{B_{\alpha\beta}}
\newcommand{\Babu}{B^{\alpha\beta}}
\newcommand{\Xa}{X^\alpha}
\newcommand{\Xb}{X^\beta}

\newcommand{\G}{\Gamma}
\newcommand{\g}{\gamma}
\newcommand{\de}{\delta}
\newcommand{\De}{\Delta}
\newcommand{\m}{\mu}
\newcommand{\n}{\nu}
\newcommand{\la}{\lambda}
\newcommand{\La}{\Lambda}
\newcommand{\s}{\sigma}
\newcommand{\Si}{\Sigma}
\newcommand{\tht}{\theta}
\newcommand{\om}{\omega}
\newcommand{\Om}{\Omega}
\newcommand{\ep}{\epsilon}
\newcommand{\vep}{\varepsilon}
\newcommand{\ve}{\varepsilon}
\newcommand{\ka}{\kappa}
\newcommand{\io}{\iota}
\newcommand{\al}{\alpha}
\newcommand{\ze}{\zeta}

\newcommand{\pd}{\partial}
\newcommand{\parens}[1]{\left(#1\right)}
\newcommand{\bracke}[1]{\left[#1\right]}
\newcommand{\pr}[1]{\left(#1\right)}
\newcommand{\br}[1]{\left[#1\right]}
\newcommand{\cbr}[1]{\left\{#1\right\}}
\newcommand{\crb}[1]{\left\{#1\right\}}
\newcommand{\norm}[1]{\left\Vert#1\right\Vert}
\newcommand{\detm}[1]{\left|#1\right|}
\newcommand{\abs}[1]{\left\vert#1\right\vert}
\newcommand{\set}[1]{\left\{#1\right\}}
\newcommand{\eps}{\varepsilon}
\newcommand{\To}{\longrightarrow}
\newcommand{\OP}[1]{\mathcal{#1}}
\newcommand{\derv}[2]{\frac{d#1}{d#2}}
\newcommand{\nderv}[3]{\frac{d^{#1}#2}{{d#3}^#1}}
\newcommand{\psig}[1]{\sigma_#1}
\newcommand{\thet}{\theta}
\newcommand{\Grf}{\widehat{\G}}
\newcommand{\Qd}[1]{Q_{(#1)}}
\newcommand{\RC}[2]{{\OP{R}^{#1}}_{#2}}
\newcommand{\RB}[2]{\overline{\OP{R}}_{#1}^{\phantom{#1}{#2}}}
\newcommand{\RTB}[2]{\OP{R}_{\bar{#1}}^{\phantom{#1}{\bar{#2}}}}
\newcommand{\RBE}[2]{e^{-i \thet_{(#1 #2)}} {r_{#1}}^{#2}}
\newcommand{\ket}{\rangle}
\newcommand{\bra}{\langle}
\newcommand{\bipd}{\stackrel{\leftrightarrow}{\partial}}
\newcommand{\pbr}[1]{\left\{#1\right\}_{\textrm{PB}}}
\newcommand{\dbr}[1]{\left\{#1\right\}_{\textrm{DB}}}
\newcommand{\deloQ}{\delta^0_Q}
\newcommand{\delIQ}{\delta^1_Q}
\newcommand{\Diff}{\textrm{Diff}}
\newcommand{\Map}{\textrm{Map}}
\newcommand{\Tr}{\textrm{Tr}\,}
\newcommand{\Det}{\textrm{Det}\,}
\newcommand{\Id}{\mathbb{I}}

\newcommand{\upket}{|\!\!\uparrow\rangle}
\newcommand{\dnket}{|\!\!\downarrow\rangle}
\newcommand{\upbra}{\langle\uparrow|}
\newcommand{\dnbra}{\langle\downarrow|}
\newcommand{\expect}[1]{\langle #1 \rangle}
\newcommand{\Psiket}{|\Psi\rangle}
\newcommand{\Psizerket}{|\Psi^{(0)}\rangle}
\newcommand{\Psioneket}{|\Psi^{(1)}\rangle}
\newcommand{\Psibra}{\langle\Psi|}
\newcommand{\Psizerbra}{\langle\Psi^{(0)}|}
\newcommand{\Psionebra}{\langle\Psi^{(1)}|}
\newcommand{\Psitwobra}{\langle\Psi^{(2)}|}
\newcommand{\Psitwoket}{|\Psi^{(2)}\rangle}

\newcommand{\Asc}{\mathscr{A}}
\newcommand{\Bsc}{\mathscr{B}}
\newcommand{\Csc}{\mathscr{C}}
\newcommand{\Dsc}{\mathscr{D}}
\newcommand{\Esc}{\mathscr{E}}
\newcommand{\Hsc}{\mathscr{H}}
\newcommand{\Lsc}{\mathscr{L}}
\newcommand{\Msc}{\mathscr{M}}
\newcommand{\Nsc}{\mathscr{N}}
\newcommand{\Osc}{\mathscr{O}}
\newcommand{\Psc}{\mathscr{P}}
\newcommand{\Qsc}{\mathscr{Q}}
\newcommand{\Rsc}{\mathscr{R}}

\newcommand{\Acl}{\mathcal{A}}
\newcommand{\Bcl}{\mathcal{B}}
\newcommand{\Ccl}{\mathcal{C}}
\newcommand{\Dcl}{\mathcal{D}}
\newcommand{\Ecl}{\mathcal{E}}
\newcommand{\Hcl}{\mathcal{H}}
\newcommand{\Mcl}{\mathcal{M}}
\newcommand{\Ncl}{\mathcal{N}}
\newcommand{\Ocl}{\mathcal{O}}
\newcommand{\Pcl}{\mathcal{P}}
\newcommand{\Qcl}{\mathcal{Q}}
\newcommand{\Rcl}{\mathcal{R}}

\newcommand{\eb}{\mathbf{e}}
\newcommand{\gb}{\mathbf{g}}
\newcommand{\hb}{\mathbf{h}}
\newcommand{\kb}{\mathbf{k}}

\newcommand{\nhat}{\hat{n}}
\newcommand{\uhat}{\hat{u}}
\newcommand{\vhat}{\hat{v}}
\newcommand{\xhat}{\hat{x}}
\newcommand{\yhat}{\hat{y}}
\newcommand{\zhat}{\hat{z}}
\newcommand{\Bhat}{\hat{B}}

\newcommand{\omhat}{\hat{\omega}}
\newcommand{\phihat}{\hat{\phi}}
\newcommand{\xihat}{\hat{\xi}}

\newcommand{\nab}{\nabla}

\newcommand{\eff}{\textrm{eff}}

\newcommand{\Ord}{\mathscr{O}}

\newcommand{\overrt}[1]{\frac{1}{\sqrt{#1}}}

\newtheorem{point}{Point}

\begin{flushright}
\break
hep-th/0512250\\
DAMTP-2005-130\\[3mm]
\end{flushright}
\vspace{1cm}
\begin{center}
{\Large {\bf Conformal Sigma-Models on Supercoset Targets}}

\vspace{0.8cm} {\large David Kagan$^a$
                       and Charles A. S. Young$^b$}\\[3mm]
{\em $^a$DAMTP, Centre for Mathematical Sciences, University of Cambridge,\\
Wilberforce Road, Cambridge CB3 0WA, UK}\\[3mm]
{\em $^b$Department of Mathematics, University of York,\\
Heslington Lane, York YO10 5DD, UK}\\
{\small E-mail: {\tt d.kagan@damtp.cam.ac.uk, charlesyoung@cantab.net}}
\end{center}

\vskip 0.15in
\centerline{\small\bf ABSTRACT}
\centerline{
\parbox[t]{5in}{\small
\noindent We investigate the quantum behaviour of
sigma models on coset superspaces $G/H$ defined by $\mathbb Z_{2n}$
gradings of $G$. We find that, whenever $G$ has vanishing Killing
form, there is a choice of WZ term which renders the model quantum conformal, at least to one loop.
The choice coincides with that for which the model is known to be classically integrable. This generalizes results
for models associated to $\mathbb Z_4$ gradings, including IIB superstrings in $AdS_5\times S^5$.}}

\vspace{1cm}

\section{Introduction and overview}
Non-linear sigma models with supermanifolds as targets are of importance both in string theory and condensed matter physics.
In condensed matter they are relevant in a variety of applications \cite{condmat}, notably the theory of the quantum Hall effect 
\cite{QHE}. There are interesting connections to Lagrangian formulations of logarithmic conformal field theories \cite{Schomerus:2005bf}. 

In string theory, the proposal of the AdS/CFT correspondence \cite{AdSCFT} prompted renewed interest in superstrings on curved Ramond-Ramond backgrounds\footnote{For more on the technical issues of formulating string theories on supermanifold targets see \cite{Grassi}.}. The Green-Schwarz action for IIB strings in $AdS_5\times S^5$, given by Metsaev and Tseytlin in \cite{MT} (see also \cite{RS}), takes the form of a sigma model on the coset superspace $PSU(2,2|4)/SO(1,4)\times SO(5)$. 
To gain insight into how this theory might be quantized, sigma models on simpler targets, the supergroups $PSL(n|n)$, were considered in \cite{BZV, BVW} (see also \cite{deBS}). As we recall below, these supergroups have vanishing Killing form and are therefore Ricci-flat, which guarantees that sigma models on them are conformal to one-loop; the result of Bershadsky, Zhukov and Vaintrob in \cite{BZV} is that they are exactly conformal. (The representation theory of the relevant superalgebras \cite{GQS} thus becomes important.) This remarkable fact should be contrasted with the more familiar sigma models on bosonic groups, where a correctly normalized WZ term (which may be thought of as a parallelizing torsion \cite{Mukhi}) must be added to the action if the theory is to be quantum conformal \cite{WZW}. 

It was subsequently shown by Berkovits, Bershadsky, Hauer, Zhukov and Zwiebach \cite{BBHZZ} that sigma models on certain \emph{quotients} $G/H$ of Ricci-flat supergroups $G$ by bosonic subgroups are again conformal to one loop --- and can be expected to be so exactly --- given a suitable WZ term. These targets include the coset superspace $PSU(2,2|4)/SO(1,4)\times SO(5)$, as well as $PSU(1,1|2)/U(1)\times U(1)$, whose bosonic geometry is $AdS_2\times S^2$. The vital property in each case is that the isotropy subgroup $H$ is the fixed point set of a $\mathbb Z_4$ grading of $G$. This, in particular, allows the addition of the required WZ term.

What is striking is that the $\mathbb Z_4$ grading was also a key element in the demonstration by Bena, Polchinski and Roiban \cite{BPR} that the sigma model on $PSU(2,2|4)/SO(1,4)\times SO(5)$ is classically integrable. (See also \cite{MP}, which contains references to the extensive recent work on this subject.) It was shown by one of the present authors \cite{Young} that a similar construction holds, 
more generally, for sigma models on spaces $G/H$ defined by $\mathbb Z_m$ gradings: for any $m$, the grading permits the addition of a certain preferred WZ term, and with this WZ term the equations of motion may be put into Lax form, ensuring integrability, just as 
in the $\mathbb Z_4$ case. 

Given this result, and the dual role played by the $\mathbb Z_4$-grading, it is natural 
to hope that conformal invariance is also present in sigma models whose targets are quotients of (again, Ricci-flat) supergroups $G$ by 
subgroups $H$ defined by gradings of order greater than $4$.
In this paper, we show that this is indeed the case, at least at one loop. 

The structure of this paper is as follows: in section \ref{setup} we introduce some notation and discuss 
supergroups with vanishing Killing form and related coset superspaces. We also write down the sigma model action with WZ term, 
and recall the one-loop beta-functions. In section \ref{geom} we compute the Ricci curvature of a torsion connection on a reductive 
homogeneous superspace, and then in section \ref{vbf} we demonstrate that, for the connection whose torsion is given by the preferred 
WZ term, this curvature vanishes, completing our argument. Some conclusions, and comments about the consequences of this
result, are given in section \ref{conc}.

\section{Sigma models on graded coset superspaces}\label{setup}
We shall consider non-linear sigma models on certain homogeneous superspaces. We begin by summarizing the relevant facts
about Lie supergroups and establishing some notation. For full details see \cite{Dict}.

\subsection{Supergroups}

Let $G$ be a Lie supergroup and let $\liealg g$ be the complexification of its Lie superalgebra. Write
\be\liealg g=\liealg g^{(0)} + \liealg g^{(1)}\label{deo}\ee for the decomposition of $\liealg g$ into its even and odd parts, and
pick a basis $\{t_a\}$ of $\liealg g$ consisting of the disjoint union of bases of $\liealg g^{(0)}$ and $\liealg g^{(1)}$. 
For every $t_a\in \liealg g^{(0)}$, let $|a|$ be an even number; for every $t_a \in \liealg g^{(1)}$, let $|a|$ be an odd 
number.\footnote{$|a|$ will be defined fully in (\ref{adef}) below; only $|a|$ modulo 2 is relevant here.}

We assume for simplicity that a suitable representation has been chosen and the $t_a$ are concrete matrices. 
Elements of $G$ near the identity are then of the form 
\be 1 + X^a t_a + \dots\ee 
where, for each $a$, the parameter $X^a$ is Grassmann even (i.e. a c-number) if $|a|$ is even and Grassmann odd if $|a|$ is 
odd.\footnote{These parameters will also satisfy some reality conditions, which depend on the choice of real form of $\liealg g$. 
(For example, if the real form is the real span of the $t_a$, the conditions are simply that the $X^a$ are real.)} 
Since $G$ is a group, the commutator
\be \left[ X^a t_a,Y^b t_b \right] = X^a t_a \,Y^b t_b - Y^b t_b\, X^a t_a\ee
closes onto some $Z^c t_c$, and, given the statistics of the $X^a$ and $Y^b$, this implies that $\liealg g$ closes under the 
graded commutator:
\be \left[ t_a, t_b \right]_\pm = t_a t_b - (-)^{|a||b|} t_b t_a = f^c{}_{ab}\, t_c,\label{glb}\ee
where we define also the structure constants $f^a{}_{bc}$. It follows from
\be 0= \left[ Z, \left[ X,Y\right] \right] +
       \left[ X, \left[ Y,Z\right] \right] +
       \left[ Y, \left[ Z,X\right] \right]\ee
for $X=X^a t_a$, $Y=Y^a t_a$ and $Z=Z^a t_a$ that the $f^{a}{}_{bc}$ satisfy the graded Jacobi identity:
\be 0= (-)^{|a||c|} f^e{}_{ad} f^d{}_{bc} +  (-)^{|b||a|} f^e{}_{bd} f^d{}_{ca} + (-)^{|c||b|} f^e{}_{cd} f^d{}_{ab}.\label{Jca}\ee 

The Killing form $K(X,Y)=K_{ab}X^a Y^b$ on $\liealg g$ is defined by
\be K_{ab} = (-)^{|d|} f^{d}{}_{ac} f^{c}{}_{bd}  \label{KF}\ee
and is an invariant (i.e. $K([X,Z],Y) + K(X,[Y,Z]) = 0$), graded-symmetric, bilinear form. The unusual feature of Lie superalgebras, in 
comparision with Lie algebras, is that there exist simple Lie superalgebras whose Killing forms vanish identically. As we recall in 
section \ref{geom} below, the vanishing of the Killing form implies that $G$ is Ricci-flat. This will be crucial for conformal 
invariance, so we will focus on these cases. However, even if $K_{ab}$ is zero, $\liealg g$ may possess a non-degenerate, 
invariant, graded-symmetric bilinear form. We assume that such a form exists and denote it by \be\left<\cdot,\cdot\right>\ee 
(In the examples in appendix \ref{Aexm}, $\left<\cdot,\cdot\right>$ is given by the supertrace in the defining representation, 
rather than the adjoint.)

\subsection{Graded coset superspaces}

We are concerned with models whose targets are coset superspaces $G/H$ with the special property that 
$\liealg g$ is $\mathbb Z_{2n}$-graded and $\liealg h$, the complexified Lie algebra of the isotropy subgroup $H$, is the
subspace of grade zero. That is, we suppose that $\liealg g$ may be written as 
a direct sum 
\be \liealg g = \liealg g_0 + \liealg g_1 +\dots+\liealg g_{2n-1} \label{grading}\ee
of vector subspaces (where  $\liealg g_0=\liealg h$), and that this decomposition respects the graded Lie bracket (\ref{glb}):
\be \left[ \liealg g_r, \liealg g_s \right]_\pm \subset \liealg g_{r+s \mod 2n}.\ee

We suppose further that this grading is compatible with the splitting (\ref{deo}) of $\liealg g$ into its even and odd graded subspaces,
in the sense that
\be \liealg g_{2s} \subset \liealg g^{(0)},\quad \liealg g_{2s+1} \subset \liealg g^{(1)},\quad\text{for } s=0,1\dots,n-1.\ee  
The basis $\{t_a\}$ can be chosen to be a disjoint union of bases of the $\liealg g_k$, and one can then consistently define
\be |a| = s \Longleftrightarrow t_a \in \liealg g_s \label{adef}.\ee  

It is also useful to adopt the following conventions for the naming of indices:
\bea  a,b,\dots  && \text{generators of $\liealg g$} \nn\\
      \alpha,\beta,\dots && \text{generators of $\liealg h$} \nn\\
      i,j,\dots     && \text{generators of $\liealg g \setminus \liealg h$}.\eea

Finally, we assume that the inner product $\left<\cdot,\cdot\right>$ respects the grading, in the sense that if 
$X\in \liealg g_r$ and $Y\in \liealg g_s$ then $\left< X,Y\right> = 0$ unless $r+s=0 \mod 2n$. Non-degeneracy implies that if 
$X\in \liealg g_r$ and $\left<X,Y\right>=0$ for all $Y\in\liealg g_{2n-r}$ then $X=0$.

One family of examples of such coset superspaces is given in appendix \ref{Aexm}.

\subsection{The action}

Sigma models on $G/H$ are most conveniently expressed in terms of a dynamical field $g(x^\mu)\in G$ (where $x^\mu$ are worldsheet 
coordinates, and $\eta_{\mu\nu}$ will be the worldsheet metric). We write
\be j_\mu = g^{-1} \partial_\mu g \in\liealg g\ee
for the current invariant under the global left action 
\be g\mapsto Ug,\quad U\in G\label{globalG}\ee 
of $G$ on itself. This decomposes into currents of definite grade:
\be j_\mu = j_\mu^\alpha t_\alpha +  j_\mu^i t_i\ee
and under the local right action
\be g \mapsto gh\label{localH}\quad\text{with}\quad h(x^\mu)\in H,\ee 
corresponding to the redundancy in the choice of representative $g$ from the coset $gH$, 
\begin{equation} 
j_\mu^\alpha \mapsto \pr{h^{-1} j_\mu h}^\alpha + (h^{-1} \partial_\mu h)^\alpha,\quad  
j_\mu^i \mapsto \pr{h^{-1} j_\mu h}^i.
\end{equation}

It is now possible to write down the most general local action for a sigma model with $B$-field constructed using 
$\left< \cdot,\cdot\right>$ and with the required symmetries. Global $G$ symmetry is ensured by writing the action in terms of $j$; 
local $H$ symmetry requires that it takes the form
\be S = \int d^2x  \left( p_{|i|} \eta^{\mu \nu} +q_{|i|} \eps^{\mu\nu} \right) j_\mu^i  j_\nu^j \left<t_i,t_j\right> \label{action}\ee 
where $p_s=p_{2n-s}$ and $q_s=-q_{2n-s}$ are arbitrary real coefficients.
(In this expression the summation convention applies as usual over the indices $i,j$. The point is that since we require only 
$H$-invariance we can rescale each pair $(s, 2n-s)$ of subspaces separately, both in the kinetic and WZ terms.)

However, a certain choice of the coefficients $q_s$ is privileged. 
It was shown in \cite{Young} that the sigma model becomes classically integrable if\footnote{Note that, as in that paper, we shall set $p_1=\dots=p_{n-1}=1, p_n=\half$ throughout, which is the natural choice in the sense that the kinetic term is then simply $\half (j-A)_\mu (j-A)^\mu$. In the $\mathbb Z_4$ case, this gives the sigma-model for the Berkovits, rather than Green-Schwarz, superstring. The latter has $p_1=0, p_2=\half$, which corresponds to a degenerate metric on G/H, and quantization requires kappa gauge-fixing.}
\begin{equation}{\label{qs}}
q_s = 1-{s\over n}.
\end{equation}

The goal of the present paper is to show that, for supergroups $G$ with vanishing Killing form, the same choice of $q_s$ \emph{also} 
ensures conformal invariance, at least at one loop order.
To see this, we first recall some well-known \cite{Ketov} facts about the renormalization-group behaviour of sigma models.

\subsection{Sigma model $\beta$-functions to one-loop}

We begin by quoting the $\beta$-functions for the general non-linear sigma model with $B$-field. Let $m,n,\dots$ denote coordinate-induced tangent-vector indices on $G/H$. Then the classical action is
\be S = \half \int d^2x \left( \eta_{\mu\nu} g_{mn} + \eps_{\mu\nu} B_{mn}\right) 
                                         \partial^\mu \phi^m \partial^\nu\phi^n\label{saction}\ee
where $\phi^m(x^\mu)$ is the sigma-model field, $g_{mn}$ is the metric on the target space and $B_{mn}=-B_{nm}$ can be regarded as the 
potential for the torsion 3-form $H = dB$.  Under changes in the energy scale $\mu$, these quantities are renormalized according to
$\mu \frac{\partial}{\partial\mu} g_{mn} = \beta_{(mn)}$ and $\mu \frac{\partial}{\partial\mu} B_{mn} = \beta_{[mn]}$,
with, to one-loop order \cite{Ketov},
\be \beta_{mn} = -\frac{\hbar}{2\pi} \hat R_{mn} + O(\hbar^2) \label{oneloop}\ee
where $\hat R$ is the generalized Ricci curvature of the metric connection for $g$ with torsion $H$. To connect the notation 
in (\ref{action}) with the general sigma model action (\ref{saction}), define
\be e_m^a = (g^{-1} \partial_m g)^a,\label{vielbein}\ee
so that $j^a_\mu = e_m^a \partial_\mu \phi^m$ and the metric and $B$-field in (\ref{action}) are $g_{mn} = e_m^{i} e_n^{j}\, g_{ij}$, 
$B_{mn} = e_m^{i} e_n^{j}\, B_{ij}$ with
\be  g_{ij} = \left< t_{i}, t_{j} \right>  p_{|i|} ,\quad  B_{ij} = \left< t_{i}, t_{j} \right>  q_{|i|}.\ee 

The task is now purely geometrical: we must show that the choice of Wess-Zumino (i.e $B$-field) term given by the coefficients 
$q_s$ in (\ref{qs}) renders $G/H$ Ricci-flat. In what follows it is easiest to work in the non-coordinate-induced basis 
(indices $i,j,\dots$ ) of tangent vectors on $G/H$ defined by the vielbeins $e_m^i$

\section{Ricci curvature of $G/H$}\label{geom}
\subsection{The torsion-free connection}
The identity $de+e\wedge e=0$ satisfied by the (super-)Maurer-Cartan form 
\be g^{-1}dg = e = e^a\,t_a = e^\al\,t_\al + e^i\,t_i.\ee
yields the structure equations
\begin{eqnarray}
de^i   &=& -\frac{1}{2}{f^{i}}_{jk} e^j \wedge e^k
           -           {f^i}_{j \al} e^j \wedge e^\al \\
de^\al &=& -\frac{1}{2}{f^\al}_{ij} e^i \wedge e^j
           -\frac{1}{2}{f^\al}_{\beta\g} e^\beta \wedge e^\g.\label{struct}
\end{eqnarray}
From the first of these one reads off the components of the torsion-free ($de^i + \omega^i{}_j \wedge e^j=0$) connection 1-form
\be {\om^i}_j = -\frac{1}{2}(-)^{|j||k|}{f^i}_{jk}\,e^k - {f^i}_{j\al}\,e^\al,\label{scn}\ee
whose curvature 2-form is
\bea {\Rsc^{\,i}}_j = d\omega^{\,i}{}_j +\omega^{\,i}{}_p \wedge \omega^{\,p}{}_j = 
 \frac{1}{4}\pr{(-)^{|j||p|}{f^i}_{jp}\,{f^p}_{kl} + (-)^{|k||p|+|j||l|}{f^i}_{pk}\,{f^p}_{jl}
 +               2{f^i}_{j\al}\,{f^\al}_{kl}}e^k\wedge e^l \eea
(The graded Jacobi identity (\ref{Jca}) ensures that the components proportional to $e^k\wedge e^\al$ and
$e^\beta\wedge e^\g$ vanish, as they must to maintain $H$-gauge-covariance.) The curvature tensor is then
\bea {R^i}_{jkl} =
            - \frac{1}{2}(-)^{|j||p|+|k||l|}{f^i}_{jp}\,{f^p}_{lk}
            + \frac{1}{4}(-)^{|k||p|+|j||l|}{f^i}_{pk}\,{f^p}_{jl}
            + \frac{1}{4}(-)^{|j||k|+|k||l|}{f^i}_{lp}\,{f^p}_{jk}
            + {f^i}_{j\al}\,{f^\al}_{kl}
\eea
and the Ricci tensor, which is graded-symmetric in $i\leftrightarrow j$ given the Jacobi identity, is\footnote{Note that $f^k{}_{kp}=0$
by virtue of the gradation of $\liealg g$.}
\be R_{ij} = (-)^{|k|} {R^k}_{jki} = -\frac{1}{4}(-)^{|k|}{f^k}_{ip}\,{f^p}_{jk} -  {f^\al}_{ip}\,{f^p}_{j\al}.\ee
Note in particular that the Ricci curvature of $G$ itself is proportional to the Killing form (to see this, suppose $H$ were the 
trivial subgroup). Now, from the definition (\ref{KF}), 
\be K_{ij} = 2f^{\al}{}_{ip} f^p{}_{j\al} + (-)^k f^k{}_{ip} f^p{}_{jk} \ee
and hence 
\be R_{ij}  = \frac{1}{4}(-)^{|k|}{f^k}_{ip}\,{f^p}_{jk} - \half K_{ij},\ee
so that, finally, if the Killing form vanishes then
\be R_{ij} = \frac{1}{4}(-)^{|k\,|}{f^k}_{ip}\,{f^p}_{jk}.\ee

\subsection{The torsion connection}
We now turn to the generalized Ricci tensor in (\ref{oneloop}). The connection with torsion $H=dB$ is
\be\hat{\om}^i_{\ j} = {\om^i}_j + \frac{1}{2}{H^i}_j,\ee
where ${H^i}_j = {H^i}_{kj}\,e^k$, for then
\be de^i + {\hat{\om}^i}_{\ j} \wedge e^j = H^i, \ee
where $H^i = \frac{1}{2}{H^i}_{jk}\,e^j\wedge e^k$ is the torsion 2-form. The curvature 2-form is
\bea
\hat{\Rsc}^{\,i}{}_j &=& d{\hat{\om}^i}{}_j + {\hat{\om}^i}{}_p \wedge {\hat{\om}^p}{}_j \label{clr}\\
                     &=& {\Rsc^{\,i}}_j + \frac{1}{2}d{H^i}_j + \frac{1}{2}{H^i}_p \wedge {\om^p}_j
                                                              + \frac{1}{2}{\om^i}_p \wedge {H^p}_j
                                                              + \frac{1}{4}{H^i}_p \wedge {H^p}_j. \nonumber \\
&=& {\Rsc^{\,i}}_j + \frac{1}{4}{H^i}_p \wedge {H^p}_j
 +  \pr{\pd_k H^i{}_{lj} - H^i{}_{pj}\,\om^p{}_{lk}
    -(-)^{|k||l|}H^i{}_{lp}\,\om^p{}_{jk} + \om^i{}_{pk}\,H^p{}_{lj}}e^k\wedge e^l\nn.\eea
and we write the expression in parentheses here as $\del_k H^i{}_{lj}$. Then
\be\label{GenCurvature}
\hat{R}^i_{\ jkl} = {R^i}_{jkl} + \frac{1}{4}{H^i}_{kp}{H^p}_{lj}
                                  - \frac{1}{4}(-)^{|k||l|}{H^i}_{lp}{H^p}_{kj}
                                  + \frac{1}{2}\del_k {H^i}_{lj}
                                  - \frac{1}{2}(-)^{|k||l|}\del_l {H^i}_{kj}, \ee
and the generalized Ricci tensor is
\begin{equation}
\hat{R}_{ij} = (-)^{|k|}{\hat{R}^k}_{\ jki}
            = R_{ij} + (-)^{|k|}\frac{1}{4}{H^k}_{ip}{H^p}_{jk} + (-)^{|k|}\frac{1}{2}\del_{k} {H^k}_{ij}.
\end{equation}
It now remains to show that this quantity vanishes, implying one-loop conformal invariance of the sigma model.

\section{Vanishing of the $\beta$-functions}\label{vbf}
We now return to the particular $B$-field specified by (\ref{qs}). As noted in \cite{Young}, this choice of $q_s$ leads to a 
particularly simple form for the torsion $H=dB$. Using the structure equations (\ref{struct}) one finds that
\be H_{ijk} = \begin{cases} + f_{ijk} & |i|+|j| +|k| = 2n\\ -f_{ijk} & |i|+|j| +|k| = 4n\\ 0 &\text{otherwise}\end{cases}\ee
(where the third index on $f$ is lowered using the metric $\left<\cdot,\cdot\right>$) and hence that the torsion tensor
$H^i{}_{jk}$, with indices in their natural positions, is
\be H^i{}_{jk} = \begin{cases} +f^i{}_{jk} & |j|+|k| = |i|
                            \\ -f^i{}_{jk} & |j|+|k| = 2n + |i|\\ 0& \text{otherwise}\end{cases}.\label{HHH}\ee
Now since the components $H^i{}_{jk}$ are constant ($\pd_l H^i{}_{jk}=0$) and invariant under the action of the isotropy subgroup $H$,
we have, from the definition of $\del$ in (\ref{clr}),
\bea \del_l {H^k}_{ij} &=& H^p{}_{ij}\,\om^k{}_{pl} - H^k{}_{pj}\,\om^p{}_{il} - (-)^{|i||l|}H^k{}_{ip}\,\om^p{}_{jl} \nn\\
                      &=& H^p{}_{ij}\,f^k{}_{lp} - H^k{}_{pj}\,f^p{}_{li} - (-)^{|i||l|}H^k{}_{ip}\,f^p{}_{lj}\eea
and hence
\bea (-)^{|k|}\del_k {H^k}_{ij}
&=& - (-)^{|k|}H^k{}_{pj}\,f^p{}_{ki} - (-)^{|k||i|+|k|}H^k{}_{ip}\,f^p{}_{kj} \\
&=& (-)^{|k|}\pr{{H^k}_{ip}\,{f^p}_{jk}-(-)^{|i||j|}{H^k}_{jp}\,{f^p}_{ik}}.\eea
(To reach the final line note that $f^i{}_{jk}$ and $H^i{}_{jk}$ are non-zero only when $|i|=|j| + |k| \mod 2n$.) 
The generalized Ricci curvature is therefore
\begin{equation}
\hat{R}_{ij} = \frac{1}{4}(-)^{|k\,|}\pr{{f^k}_{ip}\,{f^p}_{jk} + {H^k}_{ip}{H^p}_{jk}}
             + \frac{1}{2}(-)^{|k|}\pr{{H^k}_{ip}\,{f^p}_{jk}-(-)^{|i||j|}{H^k}_{jp}\,{f^p}_{ik}}.\label{Rhat}
\end{equation}

It follows almost immediately that the metric $\beta$-function vanishes: the graded-symmetric part of (\ref{Rhat}) is proportional to
\be (-)^{|k\,|}\pr{{f^k}_{ip}\,{f^p}_{jk} + {H^k}_{ip}{H^p}_{jk}}\ee
and if, in any given term in the summation on the right, $|j|+|k|=|p|$ then, since $|i|=2n-|j|$, $|i| + |p| = 2n+|k|$; 
conversely if $|j|+|k|=2n+ |p|$ then $|i|+|p|=|k|$. So in each term precisely one minus sign appears, and the $HH$ and $ff$ summations
cancel perfectly.

We now turn to demonstrating that the $B$-field $\beta$-function vanishes. This argument is considerably more subtle. 
The graded-antisymmetric part of (\ref{Rhat}) is
\be \hat{R}_{[ij]} = \half (-)^{|k|} \pr{{H^k}_{ip}\,{f^p}_{jk} - (-)^{|i||j|} {H^k}_{jp}\,{f^p}_{ik}}.\label{BB}\ee
We shall treat this grade by grade. Suppose $|i|=r$ and $|j|=2n-r$ with 
$r\in\{1,2\dots,2n- 1\}$. Then, writing $k_t$ for an index of the grade $t$ subspace, we find
\bea \hat{R}_{[i_rj_{2n-r}]} &=& - H^{p_1}{}_{i_r k_{2n-r+1}} f^{k_{2n-r+1}}{}_{j_{2n-r} p_1}   
                               + H^{p_2}{}_{i_r k_{2n-r+2}} f^{k_{2n-r+2}}{}_{j_{2n-r} p_2}   - \dots \nn\\
                   && \ldots - (-)^r H^{p_{r-1}}{}_{i_r k_{2n-1}} f^{k_{2n-1}}{}_{j_{2n-r} p_{r-1}}
                            - (-)^r H^{p_{r+1}}{}_{i_r k_{1}}    f^{k_{1}}{}_{j_{2n-r} p_{r+1}} + \dots \nn\\
                   && \ldots -  H^{p_{2n-1}}{}_{i_r k_{2n-r-1}}   f^{k_{2n-r-1}}{}_{j_{2n-r} p_{2n-1}}\nn\\
                  &=&   f^{p_1}{}_{i_r k_{2n-r+1}} f^{k_{2n-r+1}}{}_{j_{2n-r} p_1}   
                               - f^{p_2}{}_{i_r k_{2n-r+2}} f^{k_{2n-r+2}}{}_{j_{2n-r} p_2}   - \dots \nn\\
                   && \ldots + (-)^r f^{p_{r-1}}{}_{i_r k_{2n-1}} f^{k_{2n-1}}{}_{j_{2n-r} p_{r-1}}
                            - (-)^r f^{p_{r+1}}{}_{i_r k_{1}}    f^{k_{1}}{}_{j_{2n-r} p_{r+1}} + \dots \nn\\
                   && \ldots -  f^{p_{2n-1}}{}_{i_r k_{2n-r-1}}   f^{k_{2n-r-1}}{}_{j_{2n-r} p_{2n-1}},\label{BBB}\eea
where the factor $\half$ disappears because the definition of $H$ ensured that the two terms in (\ref{BB}) simply reinforced 
one another (or, equivalently, that the first term automatically had the necessary graded-antisymmetry).

The plan is to show that (\ref{BBB}) can be re-written as a certain sum of terms involving $f^{k_t}{}_{k_t \alpha}$. To do this we use 
the vanishing of the $ij$th component of the Killing form, which it is useful to write as a sum of
separate summations restricted to subspaces of definite grade:
\bea 0 &=& f^\alpha{}_{i_r k_{2n-r}} f^{k_{2n-r}}{}_{j_{2n-r} \alpha} 
- f^{p_1}{}_{i_r k_{2n-r+1}} f^{k_{2n-r+1}}{}_{j_{2n-r} p_1} + f^{p_2}{}_{i_r k_{2n-r+2}} f^{k_{2n-r+2}}{}_{j_{2n-r} p_2} -\dots\nn\\
&& \ldots -(-)^r  f^{p_{r-1}}{}_{i_r k_{2n-1}} f^{k_{2n-1}}{}_{j_{2n-r} p_{r-1}} 
          +(-)^r  f^{p_r}{}_{i_r k_{\alpha}} f^{k_{\alpha}}{}_{j_{2n-r} p_{r}}
          -(-)^r  f^{p_{r+1}}{}_{i_r k_{1}} f^{k_{1}}{}_{j_{2n-r} p_{r+1}} +\dots \nn\\
&& \ldots -  f^{p_{2n-1}}{}_{i_r k_{2n-r-1}} f^{k_{2n-r-1}}{}_{j_{2n-r} p_{2n-1}}.\label{KILL}\eea
We require also the graded Jacobi identity (\ref{Jca}), which, for the components of interest, takes the form
\be 0= (-)^{r|e|} f^{a}{}_{i_r b} f^{b}{}_{j_{2n-r} e} + (-)^r f^a{}_{j_{2n-r} b} f^b{}_{e i_r} + 
                (-)^{r|e|} f^a{}_{e b} f^b{}_{i_r j_{2n-r}}.\ee 
The key idea is that this, since it holds for all indices $a$ and $e$, holds in particular when $a=e$. One can sum the 
resulting identities over any set of indices $a$ one chooses. The sums relevant here are those over the (bases of the) individual 
subspaces of grades $r = 1,2,\dots,2n-2, 2n-1$:
\begin{eqnarray}
 0&=&f^{p_1}{}_{i_r k_{2n-r+1}} f^{k_{2n-r+1}}{}_{j_{2n-r} p_1} + f^{p_1}{}_{j_{2n-r} k_{r+1}} f^{k_{r+1}}{}_{p_1 i_r}
                              + f^{p_1}{}_{p_1 \alpha} f^{\alpha}{}_{i_r j_{2n-r}}\\
 0&=&f^{p_2}{}_{i_r k_{2n-r+2}} f^{k_{2n-r+2}}{}_{j_{2n-r} p_2} +(-)^r f^{p_2}{}_{j_{2n-r} k_{r+2}} f^{k_{r+2}}{}_{p_2 i_r}
                              + f^{p_2}{}_{p_2 \alpha} f^{\alpha}{}_{i_r j_{2n-r}}\nn\\
  &\vdots&\nn\\ 
0&=&f^{p_{2n-2}}{}_{i_r k_{2n-r-2}} f^{k_{2n-r-2}}{}_{j_{2n-r} p_{2n-2}} + 
                        (-)^r f^{p_{2n-2}}{}_{j_{2n-r} k_{r-2}} f^{k_{r-2}}{}_{p_{2n-2} i_r}
                              + f^{p_{2n-2}}{}_{p_{2n-2} \alpha} f^{\alpha}{}_{i_r j_{2n-r}}\nn\\
 0&=&f^{p_{2n-1}}{}_{i_r k_{2n-r-1}} f^{k_{2n-r-1}}{}_{j_{2n-r} p_{2n-1}} + 
                        f^{p_{2n-1}}{}_{j_{2n-r} k_{r-1}} f^{k_{r-1}}{}_{p_{2n-1} i_r}
                              + f^{p_{2n-1}}{}_{p_{2n-1} \alpha} f^{\alpha}{}_{i_r j_{2n-r}}.\nn\label{JAC}
\end{eqnarray}
Consider now subtracting from equation (\ref{BBB}) a linear combination (with coefficients, say, 
$a_1$, $a_2,\dots,a_{2n-2}$, $a_{2n-1}$) of these identities and a multiple $b$ of the identity (\ref{KILL}). The goal is to 
reduce the right-hand side of (\ref{BBB}) to an expression involving \emph{only} terms of the type 
$f^{p_t}{}_{p_t \alpha} f^{\alpha}{}_{i_r j_{2n-r}}$. This demand produces
a set of linear equations for the coefficients $a_r$ and $b$. First, the terms 
$f^\alpha{}_{i_r k_{2n-r}} f^{k_{2n-r}}{}_{j_{2n-r} \alpha}$ and $f^{p_r}{}_{i_r k_{\alpha}} f^{k_{\alpha}}{}_{j_{2n-r} p_{r}}$
do not appear if and only if
\be 0 = (-)^r b + a_r,\quad 0= b - (-)^r a_{2n-r}\ee
respectively. And the remaining terms --- the terms which are present in (\ref{BBB}) as it stands --- are cancelled if and only if
\begin{eqnarray}
       1 &=& a_1 - (-)^r a_{2n-r+1} - b,         \nn\\
      -1 &=& a_2 - (-)^r a_{2n-r+2} + b,         \nn\\ 
       1 &=& a_3 - (-)^r a_{2n-r+3} - b,         \nn\\
      &\vdots&                                   \nn\\
 (-)^r 1 &=& a_{r-1} - (-)^r a_{2n-1} - (-)^r b, \nn\\ 
-(-)^r 1 &=& a_{r+1} - (-)^r a_{2n+1} - (-)^r b, \nn\\
       &\vdots&                                  \nn\\
      -1 &=& a_{2n-3} - (-)^r a_{2n-r-3} - b,    \nn\\
       1 &=& a_{2n-2} - (-)^r a_{2n-r-2} + b,    \nn\\
      -1 &=& a_{2n-1} - (-)^r a_{2n-r-1} - b.
\end{eqnarray}
These equations are soluble: a\footnote{The solution is not always unique. In particular when $r=n$ the situation is 
simplified since $b=0$ and there is some freedom in the choice of the $a_i$.} solution is
\be a_t = - (-)^t \left( 1 - \frac{t}{n}\right),\quad b= 1-\frac{r}{n},\ee
and one is left, after these subtractions, with
\be f^{\alpha}_{i_rj_{2n-r}} \sum_{t=1}^{2n-1} (-)^t \left( 1 - \frac{t}{n}\right) f^{k_t}_{k_t \alpha}.\ee
Since the summation here does not depend on $r$, we have shown that
\be \hat{R}_{[ij]} = f^\alpha{}_{ij} \sum_{t=1}^{2n-1} (-)^t \left( 1 - \frac{t}{n}\right) f^{k_t}_{k_t \alpha}.\ee

But now, just as for the $\mathbb Z_4$ case in \cite{BBHZZ}, the two-form with these components is exact, and in fact zero if $H$ is
semisimple. The argument for exactness is as follows: call the sum on the right $V_\alpha$, and note that the graded Jacobi identity 
implies that $V_\alpha f^{\alpha}{}_{\beta \gamma}$ vanishes.\footnote{For each grade $t$, we have
\be f^{k_t}{}_{k_t \alpha} f^{\alpha}{}_{\beta\gamma} + f^{k_t}{}_{\beta p_t} f^{p_t}{}_{\gamma k_t} + 
f^{k_t}{}_{\gamma p_t} f^{p_t}{}_{k_t \beta} =0\ee and, on renaming the indices, it is clear that the last two terms cancel.} 
It then follows from the structure equations (\ref{struct}) that 
\begin{equation}
\hat{R}_{[ij]} e^i\wedge e^j \sim d V_\alpha \,e^\alpha
\end{equation}
and hence that the $B$-field is renormalized only by a total derivative, which has no physical effect since it does not modify 
the torsion $H=dB$. 

Furthermore, if the isotropy subgroup $H$ is semisimple then every generator $T_\alpha$ of $H$ can be expressed in terms of 
commutators of other generators and so the vanishing of $V_\alpha f^{\alpha}{}_{\beta \gamma}$ actually implies the vanishing of
$V_\alpha$. The two-form $\betab$ is then not merely exact but actually zero. 

\section{Conclusions and outlook}\label{conc}
In this paper we have calculated the generalized Ricci curvature of coset superspaces $G/H$ defined by $\mathbb Z_{2n}$ gradings of Ricci-flat supergroups. We have shown that these supercosets admit a Ricci-flat torsion connection, and that, consequently, sigma models with such spaces as targets are quantum conformal to one loop for an appropriate choice of WZ term. As in the $\mathbb Z_4$ case in \cite{BBHZZ} we expect that conformal invariance is exact, but further work is needed to establish this. 

This result provides strong evidence to support the suggestion, made in \cite{BZV}, that quotienting Ricci-flat supergroups --- specifically products of $PSL(n|n)$ --- by suitable subgroups might be a fruitful way to find new lagrangian conformal field theories. Theories of this type have been vital in the study of the gauge/string correspondence, and we hope that a more general class of such theories might be used to shed further light on this remarkable subject \cite{Poly1,Poly2}.

Another reason for interest in these backgrounds is their possible relevance to mirror symmetry. Some years ago, 
it was argued \cite{Sethi,ASchwarz,AganagicVafa} that certain Ricci-flat supergroups could profitably be regarded as Calabi-Yau supermanifolds, and that they actually arise as mirrors of rigid Calabi-Yau manifolds. This prompted work on Ricci-flat supermanifolds \cite{Zhou,RW,LRU}. Our targets are Ricci-flat only in the generalized sense, but one can speculate that there are links to be made. 

Finally, it is interesting that the same WZ term needed for classical integrability, as evidenced by the existence of one-parameter families of flat currents, is also necessary for generalized Ricci flatness of the target. One can hope that there are deeper connections to be made between Ricci-flat targets and classical integrability.

\bigskip

\bigskip
\noindent{\em Acknowledgements}
\bigskip

We thank Jonathan Evans for his helpful comments. D.K. is grateful for an NSF graduate research fellowship and to the University of Pennsylvania Physics Department, where some of this work was done. C.A.S.Y. thanks Nicolas Cramp\'e, Niall MacKay, Thomas Quella, and Andrei Babichenko for useful discussions, and gratefully acknowledges the financial support of PPARC.

\begin{appendix}
\section{Example of graded coset superspaces}\label{Aexm}
There are two classical families of simple Lie superalgebras for which the Killing form vanishes identically\cite{Dict}: 
$A(r-1|r-1)=\psl(r|r)$ and $D(r+1,r)=\mathfrak{osp}(2r+2|2r)$.
Consider the former. In its defining representation,
$\mathfrak{sl}(r|r)$ consists of matrices (``even super-matrices'') of the block
form\footnote{More precisely, this is the form of a general linear
combination of the superalgebra generators with coefficients of appropriate Grassmann grade.} 
\be M = \bmx A & X \\ Y & B \emx \ee where $A, B$ are
$r\times r$ matrices with bosonic entries and $X, Y$ are $r\times r$
matrices with fermionic entries with the
condition that the supertrace vanishes: \be\str M := \tr A - \tr B =
0.\ee The set of such matrices forms a Lie algebra under the matrix
commutator, but this algebra is not simple because the element
\be\bmx 1_{r\times r}&\\&1_{r\times r}\emx\ee is central and so
generates a one-dimensional ideal. One reaches $\psl(r|r)$ by
quotienting out by this ideal, so we can regard $\psl(r|r)$ as the
set of matrices $M$ as above but with $\tr A = \tr B = 0$, and the
Lie product as the matrix commutator composed with the obvious
projection back onto this subspace.

Finding a $\mathbb Z_{2n}$-grading of $\liealg g = \psl(r|r)$ is equivalent to finding an automorphism of order $2n$\footnote{That
is, a map $\sigma: \liealg g \rightarrow \liealg g$ such that $[\sigma M,\sigma L] = \sigma [M,L]$, $\sigma^{2n} =1$ and 
$\sigma^{k}\neq 1$ for all $k<2n$.} For, given such an automorphism, the grade $k$ subspace is the eigenspace with eigenvalue 
$e^{\pi i k/n}$, and conversely given such a grading one can define an automorphism by specifing its action on the subspaces of definite 
grade. We thus seek automorphisms $\sigma$ of $\psl(r|r)$ of even order
such that the ``even'' eigenspaces of $\sigma$ are bosonic and the
``odd'' eigenspaces are fermionic. That is to say: we want the eigenspace with eigenvalue $e^{p
\pi i/n}$ to be block on-diagonal, and the eigenspace with
eigenvalue $e^{(p-\half)\pi i/n}$ to be block off-diagonal, for
every $p\in \{1,2,\dots,n\}$.
One class of such automorphisms is as follows.
Let \be\tau: A\mapsto N^{-1} A N\label{tau}\ee be an (inner)
automorphism of $\mathfrak{sl}(r)$ of order $n$ (so $N^n=\pm 1$ and
hence, up to a similarity transformation\cite{Kac} and an
irrelevant overall phase,
\be \label{Ndef}  N = \bmx 1_{a\times a} &                            & &\\
                           & e^\frac{2\pi i}{n} 1_{b\times b} & &\\
                           &                            & \ddots & \\
                           &                            &        & e^\frac{2\pi i(n-1)}{n} 1_{c\times c} \emx\ee
where $a+b+\dots +c=r$). Consider the order $2n$ automorphism
$\sigma$ of $\psl(r|r)$ defined by 
\be \sigma: M \rightarrow g^{-1} M g\quad\text{ with }\quad g=\bmx N & \\ & e^\frac{\pi i}{n} N
\emx.\label{autoexample}\ee This sends \be A\mapsto N^{-1} A N,\quad
B\mapsto N^{-1} B N\ee \be X\mapsto e^{\pi i/n} N^{-1} X N, \quad Y
\mapsto e^{-\pi i /n} N^{-1} Y N;\ee and it is clear that the
``even'' eigenspace of $\sigma$ with eigenvalue $e^{2\pi i p/n}$
consists of supermatrices of the form \be \bmx A & \\ & B \emx \ee
where $A$ and $B$ are both eigenvectors of $\tau$ with eigenvalue
$e^{2\pi ip/n}$, while the ``odd'' eigenspace of $\sigma$ with
eigenvalue $e^{2\pi ip/n}e^{\pi i/n}$ consists of supermatrices of
the form \be \bmx & X\\ Y & \emx \ee where $X$ and $Y$ are
eigenvectors of $\tau$ with eigenvalues $e^{2\pi ip/n}$ and $e^{2\pi
i (p+1)/n}$ respectively.

(If one specializes to $n=2$ the resulting spaces $G/H$ are
$PSL(a+b|a+b)/(SL(a)\times SL(b)\times GL(1))^2$. When $a=b=1$ this is $PSL(2|2)/(GL(1))^2$ and one real form is
$PSU(1,1|2)/(U(1)^2)$, whose bosonic geometry is $AdS_2\times
S^2$.\cite{BBHZZ} Another possible class of automorphisms involve
the outer automorphism \be \bmx A&X\\Y&B\emx \mapsto \bmx -A^\top &
-Y^\top \\ X^\top & -B^\top \emx\ee of $\psl(n|n)$, where $A^\top$
is the transpose of $A$. The automorphism of $PSL(4|4)$ which
defines $AdS_5\times S^5$ is of this type.)

\end{appendix}

\end{document}